\documentstyle[11pt,moriond,epsfig]{article}

\def\Journal#1#2#3#4{{#1} {\bf #2}, #3 (#4)}

\def\NPB{{\em Nucl. Phys.} B}
\def\PLB{{\em Phys. Lett.}  B}
\def\PRL{\em Phys. Rev. Lett.}
\def\PRD{{\em Phys. Rev.} D}

\def\ADV{\em Adv. Phys.}
\def\PRB{{\em Phys. Rev.} B}
\def\IJMPA{{\em Int.\ Jour.\ Mod.\ Phys.} B}

\def\be{\begin{equation}}
\def\ee{\end{equation}}
\def\bea{\begin{eqnarray}}
\def\eea{\end{eqnarray}}

\def\normord#1{:\, #1 \! :}
\begin{document}
\title{Exploring the fractional quantum Hall effect with electron tunneling}

\author{ Eduardo Fradkin }

\address{Department of Physics, University of Illinois at
Urbana-\-Champaign, 1110 West Green Street,Urbana, IL 61801-3080}

\maketitle
\abstracts{In this talk I present a summary of recent work on tunnel junctions 
of a fractional quantum Hall fluid and an electron reservoir, a Fermi liquid.
I consider first the case of a single point contact. This is a an exactly
solvable problem from which much can be learned. I also   
discuss in some detail how these solvable junction problems can be used to
understand many aspects of the recent electron tunneling experiments into
edge states. I also give a detailed picture of the unusual behavior of these 
junctions in their strong coupling regime. A pedagogical introduction to the
theories of edge states is also included.
}
\section{Introduction}
\label{sec:intro}

The fractional quantum Hall effect (FQHE) has uncovered a number of
unique and unexpected quantum behaviors of two-dimensional electrons.
Experiments of electron tunneling from normal Fermi liquids into FQH
states are unique tools to explore the properties of the excitations of
FQH states and to understand directly the behavior of these fluids.
Because in the bulk these fluids are incompressible, {\it i.\ e.\/} all
excitations have a finite energy gap, it is very difficult to tunnel
into the bulk of these fluids. However, it is possible to tunnel
electrons into the edge states of droplets of these fluids. 

In this talk I will discuss a theoretical description of tunneling of
electrons into FQH edge states. In the first part of the talk I give a
quick review of the current understanding of the physics of the
bulk FQH state and of the chiral Luttinger picture of the edge states.
This part of the talk is pedagogical and I included it at the request
of the organizers. Much of this material is quite standard, and here 
I follow closely the hydrodynamic theory of X.\ G.\ Wen~\cite{wen-review}. 
In the second part of the
talk I will discuss work that I have done recently, in collaboration with 
Claudio Chamon and Nancy Sandler, on junctions of FQH edges with normal Fermi
liquids. Most of the results that I presented in this part of the talk were 
originally published in papers we coauthored (references [\ref{ref:junction}] and 
[\ref{ref:andreev}]).

Let us begin by considering the standard setup of the FQHE. That is we
will consider a two-dimensional electron gas (2DEG) formed in a Al As-Ga
As heterostructure in the presence of a high perpendicular magenetic
field $B$. As usual we will measure the electron density in terms of the
filling fraction $\nu$ of the Landau level defined as $\nu=N_e/N_\phi$,
where $N_e$ is the number of electrosn in the 2DEG and $N_\phi$ is the
total number of flux quanta piercing the 2DEG.
\begin{figure}[h!]
\vspace{.2cm}
\noindent
\hspace{1.5 in}
\epsfxsize=4.0in
\epsfbox{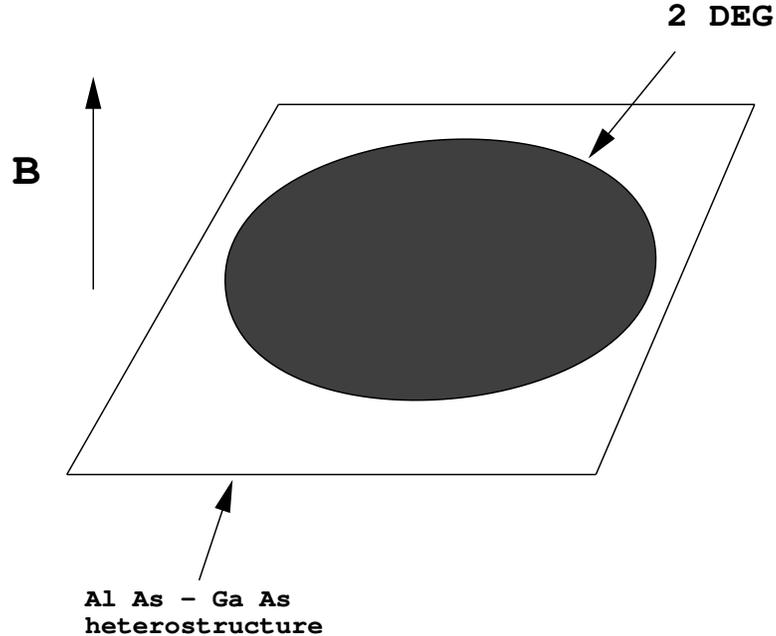}
\vspace{.5cm}
\caption{The two-dimensional electron gas in a perpendicular magnetic field.}
\label{fig:2deg}
\end{figure}
When the FQHE occurs, the 2DEG behaves as an incompressible quantum charged fluid. 
In the abscence of impurities this can only happen when certain
charge-flux commensurability conditions are met and the filling fraction
takes a set of specific (fractional) values. (Impurities trap the
excitations above these ground states in localized states,
and that is how the FQHE becomes observable.) The most prominent FQH
states seen in experiment are arranged in the sequence
\begin{equation}
\nu={\frac{p}{2np\pm 1}}
\label{eq:jain}
\end{equation}
where $p=1,2,\ldots,\infty$ and $n=0,1,2,\ldots,\infty$,
are two integers that label the states.
The $\pm $ sign refers to the FQH of electrons and to the particle-hole
reversed sequence.  Clearly $0<\nu < 1/(2n)$ for $+$, while $1/(2n) <
\nu < 1/(2n-1)$. For $p=1$ we find the states in the Laughlin sequence
$\nu=1/(2n+1)$. The sequence of states with the largest gaps has $n=1$
and it is formed by the states $\nu=p/(2p+1)=1/3,2/5,3/7,\ldots$.

Since the bulk states are fully gapped the only gapless excitations of
this fluids reside at the edge: they are deformations of the edge. Wen 
has emphasized that such
incompressible fluids are topological fluids in that, if 
placed on a  surface with non-trivial topology, their Hilbert spaces
will be sensitive only to the topology of the surface and not to the
local properties of the material.

\begin{figure}[!t]
\vspace{.2cm}
\noindent
\hspace{1.25 in}
\epsfxsize=4.0in
\epsfbox{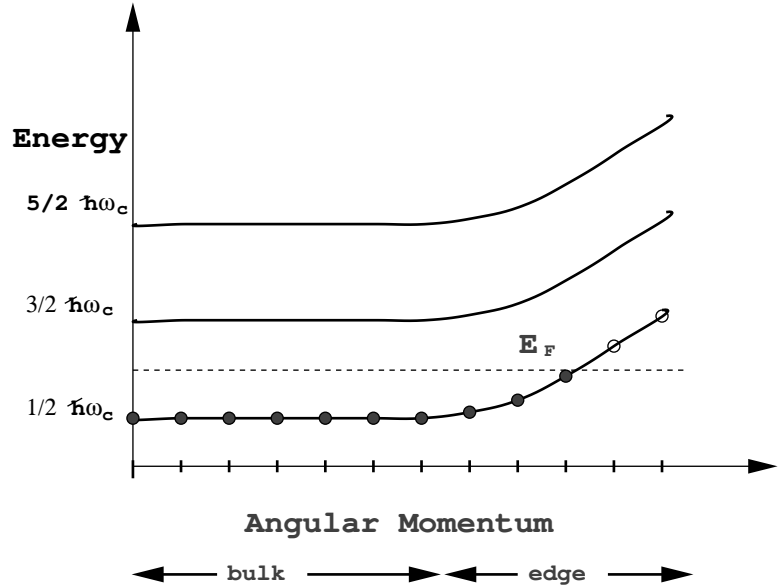}
\vspace{.5cm}
\caption{Landau levels for a non-interacting 2DEG at $\nu=1$. $E_F$ is the 
Fermi energy and $\omega_c$ is the cyclotron radius.}
\label{fig:levels}
\end{figure}

In 1982 B.\ Halperin~\cite{halperin-edge} introduced a very simple 
and appealing picture of
the edge states of non-interacting electrons in Landau levels (see
figure \ref{fig:levels}). In this picture, the bulk Landau levels ,
which have an energy spacing of $\hbar \omega_c$, are bent upwards
adiabatically near the edge of the sample by the confining potential,
with an electric field strength $E$.
For a disk-shaped sample the single particle states are eigenstates of
angular momentum (for a rotationally invariant confining potential). If
$E_F$ denotes the Fermi energy of the electrons, for a macroscopically
large sample there are many states with single particle energies in the 
vicinity of $E_F$. The level spacing of these staes is small as the 
radius of the electron droplet  gets big. In this limit the Hilbert
space of these states consists on a set of evenly spaced energy levels
with a finite density of states. These states represent electron
excitations that move in a direction specified by the sign of the magnetic 
field $B$ (or by the sign of their charge). The edge states propagate 
at the drift velocity $v=c E/B$, whre $c$ is the speed of light. 

Hence, the electrons close to the edge
behave like a chiral right moving Fermi system. In this picture, which
is a good description of a $\nu=1$ state, the only effect of the Coulomb
interactions among the electrons is a simply a change of the speed at
which the electrons at the edge propagate. If the total number of
electrons is fixed, the excitations of the edge are charge neutral
electron-hole pairs. States labelled by sets of electron-hole pairs can
also be regarded as geometric deformations of the edge of the fluid.
This picture led Wen to formulate a hydrodynamic theory of the edge
states~\cite{wen-edge}. 
\begin{figure}[h!]
\vspace{.2cm}
\noindent
\hspace{1.0 in}
\epsfxsize=4.0in
\epsfbox{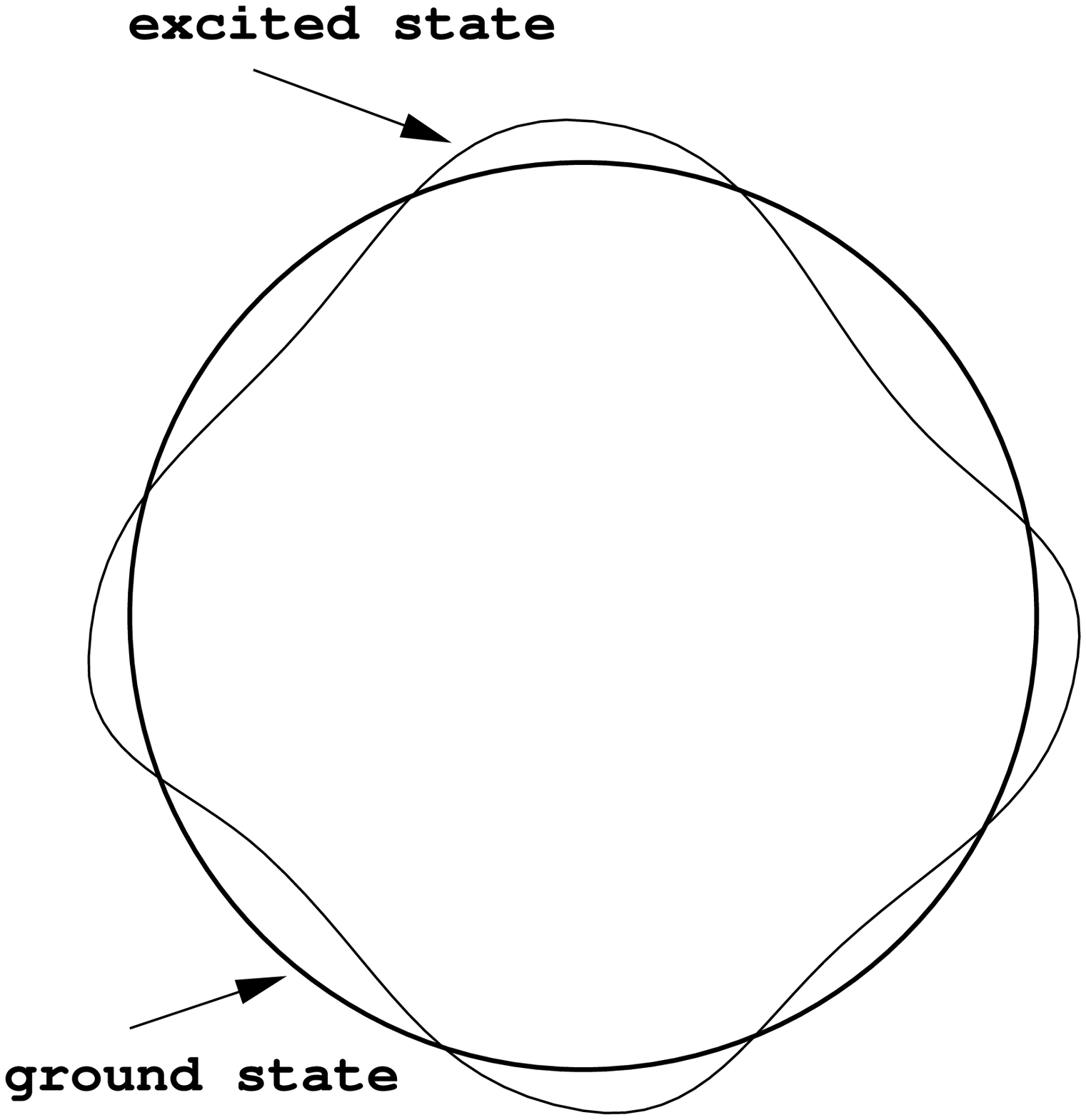}
\vspace{.5cm}
\caption{Neutral excitation of the edge.}
\label{fig:excitation}
\end{figure}
\subsection{Bulk Incompressible States}
\label{subsec:bulk}
The physics of the bulk incompressible states is by now quite well understood.
The sequence of Laughlin states $\nu=1/(2n+1)$ (with $0\leq n < \infty$)
are (almost exactly) described by the celebrated Laughlin wave
functions~\cite{laughlin}. The states in the Jain sequences can be described using 
Jain wave functions~\cite{jain}, which are generalizations of the Laughlin wave
functions. In the lowest Landau level (LLL) the single particle states
(in the circular gauge) are
\begin{equation}
\psi_r(z) \propto z^r e^{-|z|^2/(4 \ell_0^2)}
\label{eq:psi}
\end{equation}
where $\ell_0$ is the magnetic length, $\ell_0={\sqrt{e\hbar/cB}}$ and
$z$ is the (complex) coordinate of the electron, $z=x+iy$.
Let $m$ denote the odd integer $m=2n+1$. For the Laughlin states, with
filling fraction $\nu=1/m$, the Laughlin wave function for $N$
electrons in a magnetic field with $N_\phi=mN$ flux quanta is
\begin{equation}
\Psi_m(z_1,\ldots,z_N) \propto \prod_{i<j=1}^N(z_i-z_j)^m 
e^{-{\frac{1}{4\ell_0^2}}\sum_{i=1}^N|z_i|^2}
\label{eq:laughlin}
\end{equation}
 The elementary exciations in the bulk are the Laughlin quasiholes. The
wave function for a Laughlin quasihole at $z_0$ is
\begin{equation}
\Psi_m(z_0;z_1,\ldots,z_N)= \prod_{i=1}^N(z_0-z_i) \Psi_m(z_1,\ldots,z_N)
\label{eq:quasihole}
\end{equation}
Laughlin argued~\cite{Laughlin-review} that, since in this state each electron 
picks up an extra unit
of angular momentum, this state is equivalent to the effect of introducing 
(adiabatically) an infinitesimally thing solenoid which threads one flux quantum at 
$z_0$ . He further showed that the charge defect localized at $z_0$, the
quasihole, has positive charge equal to $+e/m$. As a consequence of
incompressibility there is an extra (negative) charge at the boundary
equal to $-e/m$. Hence, quasiholes make the edge swell by the right
amount to accomodate the extra charge. Furthermore, a semiclassical
argument~\cite{halperin,arovas} shows that these objects have fractional statistics
\begin{equation}
\Psi_m(z_0,z_0';z_1,\ldots,z_N)=e^{\pm i\pi/m} \Psi_m(z_0',z_0;z_1,\ldots,z_N)
\label{eq:statistics}
\end{equation}
Thus, the spectrum of excitations of these bulk incompressible states
are quasiholes with fractional charge and fractional statistics. These
results can also be (and have been) derived by field-theoretic methods
involving Chern-Simons gauge fields~\cite{zhk,lf}.

Alternatively we may also use a hydrodynamic picture of the bulk. The
hydrodynamic description is a highly economical way to summarize the
universal data of the bulk FQH states: the Hall conductance, the charge
and the statistics of the bulk excitations. It can be derived explicitly
from the Chern-Simons (mean-field) picture of the FQH states~\cite{zhk,lf} or by a
set of phenomenological arguments based on conservation laws. For the
sake of simplicity we will follow the latter approach, also introduced
by Wen~\cite{wen-review}. The fundamental idea is quite simple. The 2DEG is a charged
fluid and, as consequence of charge conservation, it can be described in
terms of the locally conserved 3-current $j_\mu$
\begin{equation}
j_\mu=\left(j_0,{\vec j}\right)
\label{eq:3current}
\end{equation}
where $j_0$ is the local charge density $\rho$, $\vec j$ is the local
charge current, and $\mu=0,x,y$ (or $\mu=0,1,2$). 
Local charge conservation means that $j_\mu$ obeys the continuity equation
\begin{equation}
\partial_\mu j^\mu=0
\label{eq:conservation}
\end{equation}
which is to say that $j_\mu$ is a locally conserved current. Since the
3-divergence of the current vanishes, it must be a curl of a vector
field $a_\mu$ (since space-time is three dimensional). Thus we write
\begin{equation}
j_\mu={\frac{1}{2\pi}}\epsilon_{\mu \nu \lambda}\partial^\nu a^\lambda
\label{eq:curl}
\end{equation}
(the factor of ${\frac{1}{2\pi}}$ is introduced for later convenience).
Observe that the 3-current is invariant if we make the local transformation 
(or redefinition) $a_\mu \to
a_\mu+\partial_\mu \Lambda$, where $\Lambda$ is an arbitrary, smooth
 single-valued function. Hence there is a gauge symmetry that is natural in this 
description. 

What do we know about the FQH state? First of all we know that the Hall conductance
has the exact value $\sigma_{xy}={\frac{\nu}{2\pi}}{\frac{e^2}{\hbar}}$.
If we denote by $A_\mu$ an external electromagnetic vector potential that we 
will use to probe the system ({\it i.\ e.\/} it is neither part of the
uniform magnetic field nor of the confining potential), the effective action
 must have a local coupling of the form $j_\mu A^\mu$. Also the
effective action must be invariant under gauge transformations of the
hydrodynamic gauge field $a_\mu$. This effective action must be local,
be odd under time reversal (and parity) and have as few derivatives as
possible. In addition it should contain only the universal data of the
FQH state which is a set of dimensionless numbers. Thus, the only terms
that can be allowed must be dimension three gauge invariant operators.
For $\nu=1/m$, the unique choice that satisfies all these constraints is 
\begin{equation}
S_{\rm eff}={\frac{m}{4\pi}}\int d^3x 
\epsilon_{\mu \nu \lambda}a^\mu \partial^\nu a^\lambda
+ {\frac{e}{2\pi}} \int d^3x 
\epsilon_{\mu \nu \lambda}A^\mu \partial^\nu a^\lambda
\label{eq:CS}
\end{equation}
where we have used eq.\ \ref{eq:curl} to write the $j_\mu A^\mu$
coupling.  It is straightforwrd to show that the current induced by
$A_\mu$ is
\begin{equation}
J^{em}_\mu=\sigma_{xy} \epsilon_{\mu \nu \lambda} \partial^\nu A^\lambda
\label{eq:jem}
\end{equation}
Hence, we get the correct value of the Hall conductance for $\nu=1/m$.

In this picture, the elementary excitations (quasiholes) look like vortices. 
Since we are interested in the low energy regime, we can assume that the 
quasiholes are quasistatic (namely, that their gap is very large) and as 
such they can be  pictured by a set of (prescribed) vortex currents
$j^\mu_{\rm vortex}$, minimally coupled to the hydrodynamic field
$a_\mu$,
\begin{equation}
S_{\rm vortex}=\int d^3x \;\; a^\mu j^\mu_{\rm vortex}
\label{eq:vortex}
\end{equation}
It is very easy to show that the exciations described by Eq.\
\ref{eq:vortex} have charge $Q=e/m$ and fractional statistics
$\theta=\pi/m$. Hence, the vortices are the Laughlin quasiholes.

\subsection{Hydrodynamic Picture of Bulk and Edge (after X.\ G.\ Wen)}
\label{subsec:hydro}

There is a very appealing picture of the dynamics of the edge
states, whcich was also introduced by Wen~\cite{wen-edge}. The theory is 
particularly simple for the case of the Laughlin states (which have filling fraction
$\nu=1/m$). In this case the edge is described in terms of a single hydrodynamic 
field. The picture of the more general FQH state is more complex and depends on 
details such as possible edge reconstructions. A good general discussion can be 
found in Wen's review ~\ref{ref:wen-review} and in my recent paper with 
Ana Lopez~\cite{ana}. Here I will consider only the edges of Laughlin states.

\begin{figure}
\vspace{.2cm}
\noindent
\hspace{1.25 in}
\epsfxsize=4.0in
\epsfbox{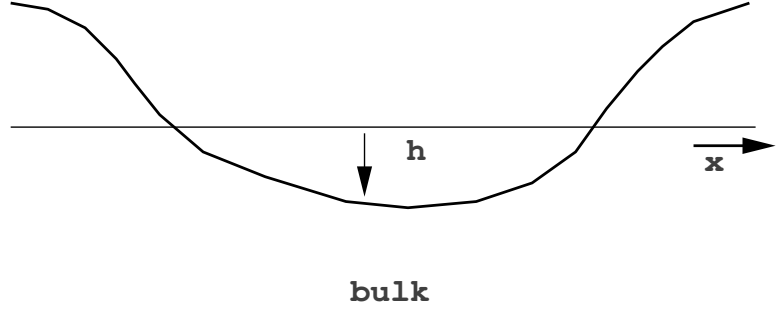}
\vspace{0.5cm}
\caption{An edge distortion.}
\label{fig:edge}
\end{figure}

Consider a droplet of  the incompressible  FQH fluid. 
For simplicity, we will picture the edge
of the fluid in its ground state as  a straight line (the horizontal
line of figure \ref{fig:edge}). In this picture, an excitation is a
deformation of the edge and in figure \ref{fig:edge} is shown as a wavy
curve. Let $h$ denote the local displacement (or height) of the fluid at
a point $x$ along the edge. Because the fluiid is incompressible, the
change of the local charge density $\delta \rho$  is proportional to the
displacement $h$, $\delta \rho \propto h$. The energy of this
excitation is the work done against the confining potential to create this 
displacement is 
\be
{\rm energy} \propto  \int dx [\delta \rho (x)]^2
\label{eq:energy}
\ee
which plays the role of a classical Hamiltonian. 
At the quantum level, the dynamics of the edge is specified by the energy 
\ref{eq:energy},
and by the commutation relation of the edge degrees of freedom. 
Since the edge degrees of freedom are chiral (namely, the move at the
drift velocity), the local density operators $\delta \rho(x)$ nave to
obey appropriate commutation relations ta lead to chiral propagation along the
edge. The (equal time) commutaion relations are
\be
\left[\delta \rho(x),\delta \rho(x')\right]=i \pi {\rm sign} (x-x')
\label{eq:ccr}
\ee
These commutation relations can be solved by introducing a filed $\phi(x)$, 
known as the chiral boson field, with the action
\be
S_{\rm edge}={\frac{1}{4\pi}} \int dx dt \partial_x \phi \left( \partial_t \phi 
- v \partial_x \phi \right)
\label{eq:action-edge}
\ee
where $v$ is the effective velocity of th edge waves. The chiral boson
is realted to the density fluctuations by
\be
\delta \rho(x)= {\frac{\sqrt{\nu}}{2\pi}} \partial_x \phi
\label{eq:relation}
\ee
With these definitions it is starightforward to show that the density
operators obey the commutation relations of Eq.\ \ref{eq:ccr} and that the 
energy of the excitaitions is given by Eq.\ \ref{eq:energy}.
In order to  have a completely well defined effective quantum theory we
need to specify the Hilbert space on which these operators act. Thus we
need to specify what type of excitations are allowed and what is their charge
and statistics. Since the bulk FQH state can only support excitations
with a charge which must be a multiple of the quasiparticle charge.
~From Eq.\ \ref{eq:relation} we find that if we add (or remove) an amount of 
 charge $Q$ to the system, the chiral boson must obey the following
boundary condition
\be
Q=\oint_{\rm edge} dx \; \delta \rho(x)={\frac{\sqrt{\nu}}{2\pi}}
\left[\phi(L)-\phi(0)\right]
\label{eq:bcs}
\ee
where the integral runs over a (closed) edge of length $L$. Here we have
set the electric charge to unity. Hence, since in the ground state sector
$Q=0$, the chiral boson obeys periodic boundary conditions (PBCs),
\be
\phi(x)=\phi(x+L)
\label{eq:pbcs}
\ee
Next we define an operator $\psi_e(x)$ which creates an {\it electron},
namely a {\it fermion} with charge $1$. In terms of the chiral boson,
the electron operator is given by the vertex operator
\be
\psi_e(x) \propto e^{{\frac{i}{\sqrt{\nu}}} \phi(x)}
\label{eq:electron}
\ee
The electron propagator is (with an $i\epsilon$ regulator implied)
\be
\langle T \psi^\dagger_e(x,t)  \psi_e(x',t') \rangle \sim
{\frac{1}{(z-z')^m}}
\label{eq:eprop}
\ee
which has a pole of order $m$. In Eq.\ \ref{eq:eprop} $T$ stands 
for time ordering and  $z=x-vt$. 

Likewise, the quasiparticle operator $\psi_{qp}(x)$ is given by 
\be
\psi_{qp}(x) \propto  e^{i\sqrt{\nu} \phi(x)}
\label{eq:qp}
\ee
which has the propagator
\be
\langle T \psi^\dagger_{qp}(x,t)  \psi_{qp}e(x',t') \rangle \sim
{\frac{1}{(z-z')^{1/m}}}
\label{eq:qpprop}
\ee
The power $1/m$ in this  propagator implies that it has a branch cut
(instead of a pole) which says that the statistics of the quasiparticle
is $\pi/m$. 

We now notice that both the electron and the quasiparticle oparator anre
invariant under the transformation $\phi \to \phi + 2 \pi n R$ where $n$
is an arbitrary integer and the compactification radius (or Luttinger parameter) 
$R=\sqrt{\nu}$. Hence, the addition of $n$ quasiparticles of charge $Q=n \nu=n/m$
implies that the chiral boson acquires the twisted boundary condition
\be
\phi(x+L)-\phi(x)={\frac{2\pi}{\sqrt{\nu}}}Q=2\pi nR
\label{eq:tbcs}
\ee
However, it is easy to see that the electron operator remains single valued 
\be
\psi_e(x+L)=\psi_e(x)
\label{eq:ebcs}
\ee
as it is required by the single-valuedness of its wave functions, while
 the quasiparticle operator instead obeys twisted boundary conditions,
\be
\psi_{qp}(x+L)=e^{i2\pi n \nu} \psi_{qp}(x)
\label{eq:qpbcs}
\ee
and are multi-valued, as requierd by the branch cut in Eq.\
\ref{eq:qpprop}.

The set of conditions that we have discussed in this section
constitute a complete definition
of the Hilbert space of the edge states. In the second part of this talk
I will use this machinery to explore a number of interesting physical
questions found in the problem of tunneling of
electrons into FQH edge states.

\section{Model of a Quantum Point Contact Junction}
\label{model}

\begin{figure}[h!]
\vspace{.2cm}
\noindent
\hspace{1.25 in}
\epsfxsize=3.0in
\epsfbox{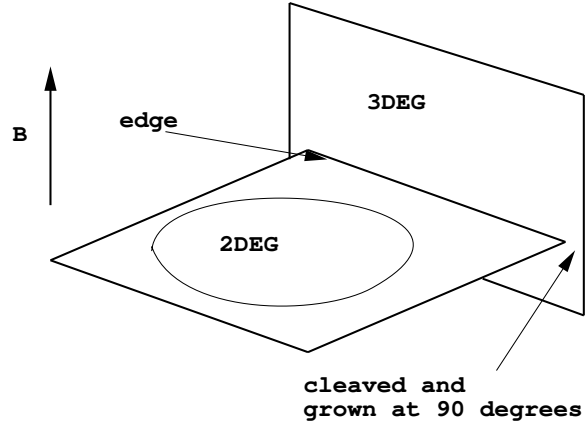}
\vspace{.5cm}
\caption{Schematic geometry of the experimental setup of A.\ Chang
and coworkers.}
\label{fig:broad}
\end{figure}

\begin{figure}[h!]
\vspace{.2cm}
\noindent
\hspace{1.5 in}
\epsfxsize=3.0in
\epsfbox{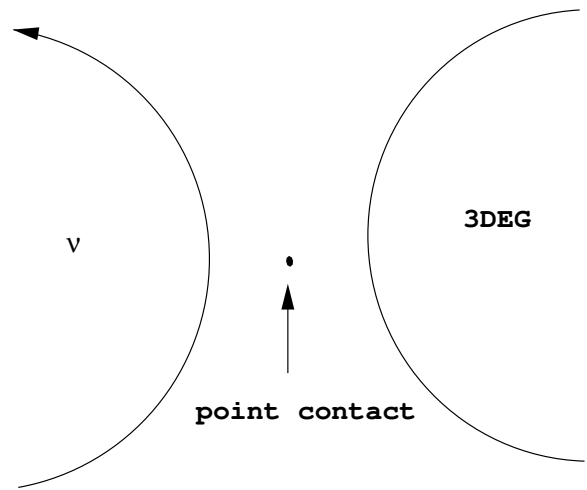}
\vspace{.5cm}
\caption{A quantum point contact between a 3DEG and the edge of a 
Laughlin fluid.}
\label{fig:contact}
\end{figure}

Let us consider first the problem of a quantum point contact junction
between an electron reservoir and the edge of a
Laughlin FQH state. Typically the electron reservoir is a
three-dimensional electron gas (3DEG). In particular, we will have in
mind the situation of the tunnel junctions of the exepriments of A.\
Chang and coworkers~\cite{chang}, see figure \ref{fig:broad}. 
In this experiment, the 3DEG acts as a reservoir of electrons. A
3DEG is well described by Fermi liquid theory. It is believed that 
the edge that results from this device is ``atomically sharp". If so,
presumably there is no edge reconstruction taking place and the electron
density of the 2DEG falls off rader rapidly and monotonically close
enough to the edge.
However, it is unclear to what degree these assumption do hold and, in
fact, in ref.\ \ref{ref:shytov} it is argued that there is a substantial
amount of charge redistribution close to the edge. In any event, even if
this were the case, it is most likely that tunnel of electrons from the
3DEG reservoir to the edge of the FQH state in the 2DEG will take place
at special places where the tunneling amplitude is largest. In other
words, the broad edge can be reagrded as a (large) set of weak tunneling
centers. In ref.\ \ref{ref:junction} C.\ Chamon and I developed a
theory of tunneling that applies to this regime. Thus, I will concentrate 
first on the description of a single tunneling center, a quantum point 
contact (QPC), depicted in figure \ref{fig:contact}. 

For the case of a single QPC the tunneling problem is substantially
simpler to describe. The theory of the QPC is highly reminiscent of the
physics of quantum impurities in metals. In fact, for a QPC, the problem 
reduces to a quantum impurity (or rather a tunneling problem) in a
one-dimensional strongly correlated (chiral) system. In particular, it
is very simple to see that,just as in the case of impurities in metals,
of the infinitely many degrees of
freedom of the 3DEG, only one channel is able to tunnel. The difference
is that while in the case of the quantum impurity in the bulk of
a metal, only the $S$-wave channel scatters off the impurity. In contrast,
in the case of the the semi-infinite geometry of the 
QPC only the electrons in the $\ell=1$ $m=0$ channel are able to tunnel. 
Nevertheless, what matters is that there is one and only one channel 
needed to describe a QPC. Details of the properties of the states
in the channel are absorbed in the definition of the tunnel matrix
element $\Gamma$.

Thus, in this picture, the 3DEG reduces to a semi-infinite 
one-dimensional Fermi liquid.
In the limit of zero tunneling matrix element, $\Gamma \to 0$, the total
charge current at the ``end point" at $x=0$ vanishes exactly. The states
of a Fermi liquid are essentially equivalent to those of a free Fermi
system, up to a renormalization of the Fermi velocitie (screening
effects and othere renormalizations of the coupling constants of the
3DEG in the bulk play almost no role in the present situation
and can be ignored.) 
\begin{figure}[h!]
\vspace{.2cm}
\noindent
\hspace{1.25 in}
\epsfxsize=4.0in
\epsfbox{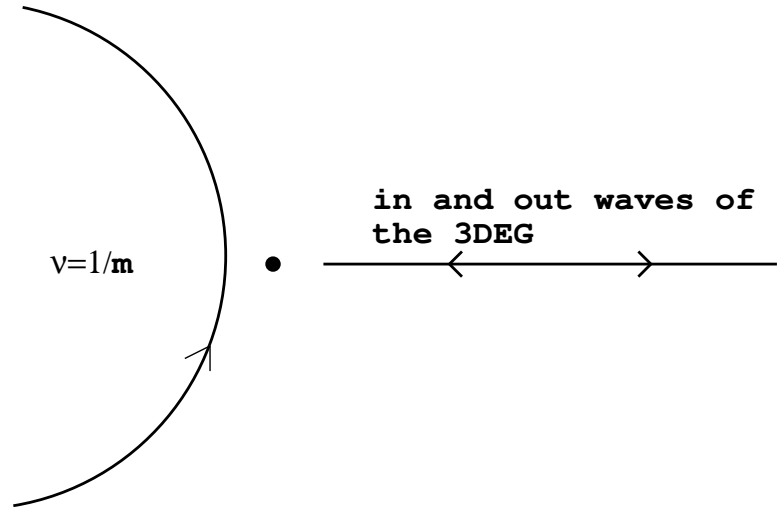}
\vspace{.5cm}
\caption{Only one channel of the 3DEG couples at the quantum point
 contact.}
\label{fig:waves}
\end{figure}

The electron states on the half-line are incoming 
and outgoing waves, as shown in figure \ref{fig:waves} satifying 
zero-current boundary conditions at the endpoint. However, exactly as in
the problem of magnetic impurities in metals, this system
exactly equivalent to a system of free chiral fermions on a full line.
Thus, the states of the electrons that participate in tunneling
processes are described by a theory of free chiral fermions. This
also happens to be the theory of the edge states of a 2DEG at $\nu=1$.
Therefore the QPC is equivalent to a junction between the edge of
a  $\nu=1$ QH state and the edge of a $\nu=1/m$ FQH state~\cite{junction}.

\begin{figure}[h!]
\vspace{.2cm}
\noindent
\hspace{1.5 in}
\epsfxsize=3.0in
\epsfbox{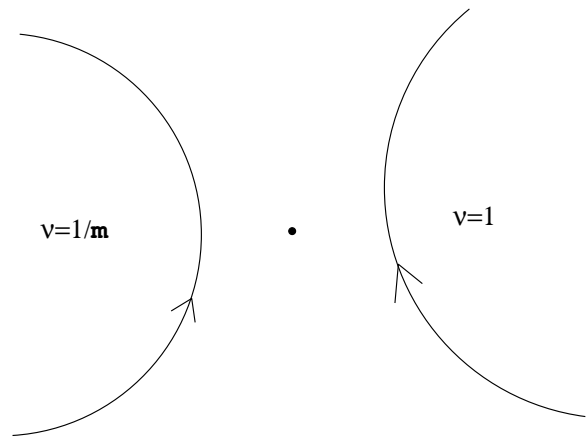}
\vspace{.5cm}
\caption{The QPC between a 3DEG and a $\nu=1/m$ FQH edge is equivalent to 
a junction between the same FQH edge and a $\nu=1$ edge.}
\label{fig:equivalent}
\end{figure}

We can now use the picture developed in the first part of the talk~\cite{junction},
section \ref{sec:intro} to write down an effective Lagrangian for the
theory of a QPC between a $\nu=1$ state (an electron resrvoir: a
3DEG or any other Fermi liquid) and the edge of a $\nu=1/m$ Laughlin FQH state.
To this end let us introduce to chiral bosons, $\phi_1$ and $\phi_2$ which
I will use to represent the Hilbert space of the edges of the FQh state and 
the Fermi liquid respectiveliy. The total Lagrangian has the form
\be
{\cal L}={\cal L}_{\rm FQH}+{\cal L}_{\rm reservoir}+{\cal L}_{\rm
tunnel}
\label{eq:lagrangian1}
\ee
where ${\cal L}_{\rm FQH}$ is the Lagrangian of the edge states of a
$\nu=1/m$ FQH fluid,
\be
{\cal L}_{\rm FQH}={\frac{1}{4\pi}} \partial_x \phi_1 \left( \partial_t \phi_1 
-  \partial_x \phi_1 \right)
\label{eq:lagrangianfqh}
\ee
and ${\cal L}_{\rm reservoir}$ is the Lagrangian for the electrons in the
reservoir,
\be
{\cal L}_{\rm reservoir}={\frac{1}{4\pi}} \partial_x \phi_2 \left
( \partial_t \phi_2 -  \partial_x \phi_2 \right)
\label{eq:lagrangianreservoir}
\ee
where we have set the velocities of both systems to unity $v_1=v_2=1$.
This is justified for a single QPC.

Finally, ${\cal L}_{\rm tunnel}$ represents the tunneling ef electrons,
\be
{\cal L}_{\rm tunnel}=\Gamma \;\delta(x) \; e^{i \omega_0 t}\; 
\normord{e^{{\frac{i}{\sqrt{\nu}}} \phi_1}}
\normord{e^{-i\phi_2}}+ {\rm h.\ c.\ }
\label{eq:tunnel}
\ee
Here, $\Gamma$ denotes the tunneling matrix element for electrons,
$\omega_0=eV/\hbar$ is the  ``Josephson" frequency (for a voltage drop of
$V$, which we will set to zero for the most part of our discussion), and 
$\exp{{\frac{i}{\sqrt{\nu}}} \phi_1}$ and $\exp{-i\phi_2}$ are the
operators that create an electron at the FQH edge and destroy an
electron at the reservoir respectively.

In reference \ref{ref:junction} it was shown that this problem can be
solved exactly. Here I will give a summary of that solution. 
The first step is to observe that by an unitary (actually orthogonal)
transfromation of the form
\be
\left(
\begin{array}{c}
\varphi_1\\
\varphi_2
\end{array}
\right)
=
\left(
\begin{array}{cc}
 \cos \theta & \sin \theta \\
-\sin \theta & \cos \theta
\end{array}
\right)
\;
\left(
\begin{array}{c}
\phi_1 \\
\phi_2
\end{array}
\right)
\label{eq:orthogonal}
\ee
the Lagrangian can be brought to the form
\be
{\cal L}=
{\frac{1}{4\pi}} \partial_x \varphi_1 \left
( \partial_t \varphi_1 -  \partial_x \varphi_1 \right)+
{\frac{1}{4\pi}} \partial_x \varphi_2 \left
( \partial_t \varphi_2 -  \partial_x \varphi_2 \right)
+
\Gamma \; \delta (x) \; 
e^{{\frac{i}{\sqrt{g'}}}( \varphi_1-\varphi_2)}
+ {\rm h.\ c.\ }
\label{eq:Ltotal}
\ee
where we have chosen the angle $\theta$ to be the solution of
\be
\cos 2 \theta= {\frac{2 \sqrt{\nu}}{1+\nu}}
\label{eq:theta}
\ee
and the ``effective filling factor" $g'$ is
\be
g'^{-1}={\frac{1}{2}} (1+\nu^{-1})
\label{eq:g'}
\ee
In other terms, we have mapped the junction between a $\nu=1$ edge and
another edge of a FQH stae at filling factor $\nu$ to a problem of
tunneling of chrage $1$ particles between two identical states at
filling factor $\nu'=g'$. Hence, the new (rotated) chiral bosons have
compactification radius $R={\sqrt{g'}}$. For the special case of $\nu=1/3$ we
find $\nu'=g'=1/2$. Notice that, the statistics of the effective 
charge $1$ particles that tunnel between the two equivalent edges is
$\theta_{\rm eff}={\frac{\pi}{g'}}$, and
for the particular case of $\nu=1/3$ they are bosons. 

Finally, we define the orthogonal fields $\varphi_\pm$
\be
\varphi_\pm={\frac{1}{\sqrt{2}}} (\varphi_1 \pm \varphi_2)
\label{eq:pm}
\ee
in terms of which the Lagrangian decouples into two pieces, ${\cal L}_+$ and 
${\cal L}_-$. In Eq.\ \ref{eq:pm} ${\cal L}_+$ is the Lagrangian of a free chiral 
boson
\be
{\cal L}_+={\frac{1}{4\pi}} \partial_x \varphi_+ \left( 
\partial_t \varphi_+ -  \partial_x \varphi_+ \right)
\label{eq:L+}
\ee
with compactification radius $R_+={\sqrt{g'/2}}$. The field $\varphi_+$
represents  the fluctuations of the conserved charge current of the
combined system. 

The Lagrangian ${\cal L}_-$ 
contains the physics of the tunneling processes. It has the form
\be
{\cal L}_-={\frac{1}{4\pi}} \partial_x \varphi_- \left( 
\partial_t \varphi_- -  \partial_x \varphi_- \right)+
\Gamma \; \delta (x) \; e^{i {\sqrt{{\frac{2}{g'}}}}\varphi_-}
+{\rm h.\ c.\ }
\label{eq:L-}
\ee
The compactification radius $R_-$ of the field $\varphi_-$ is also 
$R_-={\sqrt{g'/2}}$.

It is useful to introduce now an alternative equivalent form of the
Lagrangian of Eq.\ \ref{eq:L-}, in which instead of working with the
chiral bose field $\varphi_-$, defined on a full line of length $L$ , one defines a
{\sl non-chiral} bose field $\varphi_-$ on a half-line, $0\leq x < L/2$.
The chiral field is identified with the right and left moving components 
$\varphi_R$ and $\varphi_L$ as follows
\begin{equation}
\begin{array}{ccc}
\varphi_-(x)= &\varphi_-^R(x), & {\rm for}\;\; x>0 
\\
\varphi_-(x)= &\varphi_-^L(-x),& {\rm for}\;\; x<0 
\\
\varphi_-(x)=&\varphi_-^R(x)+\varphi_-^L(x) , &{\rm for}\;\; x\geq 0
\end{array}
\label{eq:folding}
\end{equation}
Hence, we have folded the line and we now have a non-chiral bose field
on a half-line. This procedure is very common for solving quantum impurity
problems~\cite{NFLL}.

The action for the field $\varphi_-$ on the half-line is
\be
S[\varphi_-]={\frac{1}{8\pi}}\int_0^{L/2} \int_{-\infty}^{+\infty}dt
\left[
\left({\frac{\partial \varphi_-}{\partial t}}\right)^2 -
\left({\frac{\partial \varphi_-}{\partial x}}\right)^2
\right]
+\int_{-\infty}^{+\infty}dt \;\; 2 \Gamma \; \cos \left({\frac{1}{\sqrt{2g'}}}
\varphi_-(0,t)\right)
\label{eq:BSG}
\ee
where the field $\varphi_-$ obeys Neumann boundary conditions both at $x=0$
and at $x \to \infty$. Notice that the compactification radius of
$\varphi_-$ is $R={\sqrt{2g'}}$.
\begin{figure}[h!]
\vspace{.2cm}
\noindent
\hspace{1.5 in}
\epsfxsize=3.0in
\epsfbox{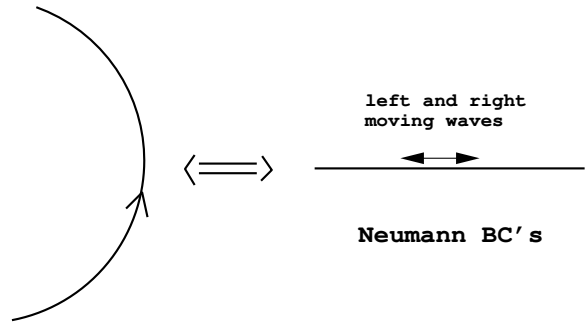}
\vspace{.5cm}
\caption{Folding a chiral boson on a non-chiral boson on a half-line with Neumann
boundary conditions.}
\label{fig:fold}
\end{figure}
The Lagrangian of Eq.\ \ref{eq:BSG} is known as the boundary sine-Gordon theory. It is
an integrable quantum field theory and a lot of useful
information is known about it~\cite{GS}. 

\begin{figure}[h!]
\vspace{.2cm}
\noindent
\hspace{1.5 in}
\epsfxsize=3.0in
\epsfbox{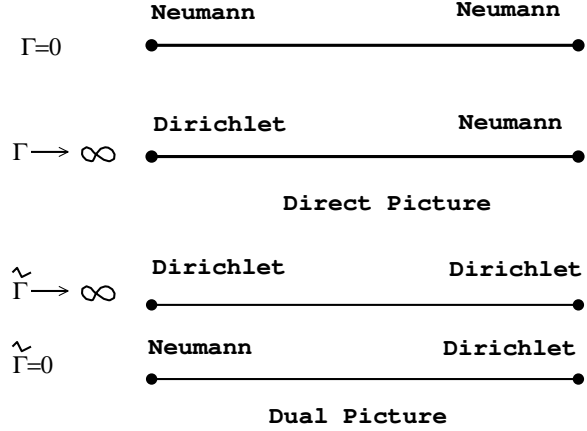}
\vspace{.5cm}
\caption{Duality as a change of boundary conditions.}
\label{fig:duality}
\end{figure}

Hence, upon folding, the problem that we need to understand is a
boundary sine-Gordon field theory with Neumann (N) boundary conditions
at both ends of the line, $\partial_x \varphi_-(x,t)|_{0,L/2}=0$. 
By inspection of the Lagrangian of Eq.\ \ref{eq:BSG} we
can see that the effect of the tunneling operator is a change in the
boundary condition (BC) at $x=0$. Indeed, for $\Gamma=0$ we have a Neumann
BC at $x=0$, whereas for $\Gamma \to \infty$ we have a Dirichlet (D) BC
at $x=0$, $\varphi_-(0,t)=0$ (see top part of fig.\ \ref{fig:duality}).
Finally we make a {\sl duality transformation} on this theory. The duality
transformation~\cite{duality}~\cite{Tduality} exchanges oscillators with kinks (or
solitons). In the folded picture it is equivalent to a replecement of
the field $\varphi_-$ by its Cauchy-Riemann dual ${\tilde \varphi}_-$,
\begin{eqnarray}
\partial_x {\tilde \varphi}_-=&& \;\; \; \partial_t \varphi_-
\nonumber \\
\partial_t {\tilde \varphi}_-=&&- \partial_x \varphi_-
\label{eq:cauchy}
\end{eqnarray}
It is straightforward to see that if the field $\varphi_-$ obeys a Neumann
(Dirichlet) BC the dual field ${\tilde \varphi}_-$ obeys a Dirichlet
(Neumann) BC. 

In addition to a change of boundary conditions, under duality the
compactification radius $R_-={\sqrt{2g'}}$ of the non-chiral field $\varphi_-$
becomes~\cite{duality,Tduality}
\be
{\tilde R}_-={\frac{2}{R_-}}=\sqrt{\frac{2}{g'}}
\label{eq:Rdual}
\ee
Hence, the dual value of $g$ is ${\tilde g}=1/g'$. 
Likewise~\cite{junction,duality}, the dual
coupling constant becomes 
\be
{\tilde \Gamma} \propto \Gamma^{-g'}
\label{eq:gammadual}
\ee
and duality maps strong coupling to weak coupling and viceversa. The
work of Fendley, Salaur and Warner \cite{FSW} has confirmed the non-perturbative
validity of these duality results.

\begin{figure}[h!]
\vspace{.2cm}
\noindent
\hspace{1.5 in}
\epsfxsize=3.0in
\epsfbox{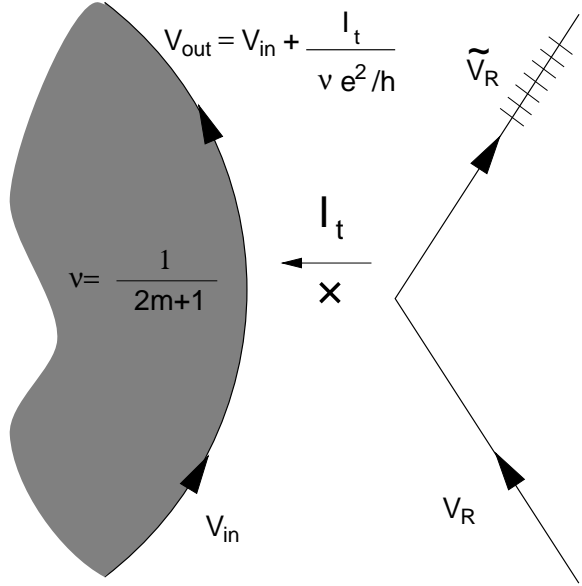}
\vspace{.5cm}
\caption{Tunnel junction and the definition of the tunneling current.}
\label{fig:tunnel}
\end{figure}

We are now ready to discuss the properties of the QPC. Here I will
follow closely the results of my work with Chamon \cite{unction} which, in
turn, relied heavily on earlier results of Fendley, Ludwig ans Saleur~\cite{FLS}.
In particular, it is possible to compute exctly the differential
tunneling
conductance $G_t={\frac{dI}{dV}}$ of the QPC, where $V$ is the voltage
drop between the FQH edge and the reservoir, $V=V_R-V_{in}$ (see figure
\ref{fig:tunnel}).
The boundary sine-Gordon theory is a free scalar field coupled to 
the vertex operator ${\cal O}=\exp({\frac{i}{\sqrt{2g'}}}
\varphi_-(0,t))$ at the boundary. The (boundary) scaling dimension of
this operator, at the weak coupling fixed point $\Gamma \to 0$, is 
$d_{\cal O}=g'^{-1}=(m+1)/2$. Thus, for $\nu=1/3$, this is operator has
boundary dimension $2>1$ and it is an irrelevant operator. Therefore,
the weak coupling fixed point is stable. 
The effective coupling vanishes as the energy scale$V$ (the voltage) is
lowered,
and all quantities, such as the conductance, that depend on $\Gamma$ scale to zero
in this regime. 

Conversely, the strong coupling fixed point is
(infrared) unstable. In this regime, $\Gamma \to \infty$ and ${\tilde
\Gamma}\to 0$. The (boundary) scaling dimension of this operator in the
dual theory (at the fixed point ${\tilde \Gamma}\to 0$), ${\tilde {\cal
O}}=\exp(i{\sqrt{2g'}} {\tilde \varphi}_-(0,t))$, is $d_{\tilde
{\cal O}}=g'=2/(m+1)<1$, and it is always relevant. Thus, for $\nu=1/3$, the 
scaling dimension of this operator is ${\frac{1}{2}}$.

By using the results of Fendley, Ludwig ans Saleur~\cite{FLS}, Chamon
and I  calculated the differential tunnel conductance $G_t$ and
found it to be \cite{junction}
\be
G_t={\tilde g}\ \frac{e^2}{h}
\times
\cases{
\sum_{n=1}^{\infty}c_n(1/{\tilde g})
\ \left(\frac{V}{2T_K}\right)^{2n(1/{\tilde g}-1)},
&\mbox{$\frac{V}{2T_K}< e^{\delta}$}\cr
1-\sum_{n=1}^{\infty}c_n({\tilde g})
\ \left(\frac{V}{2T_K}\right)^{2n({\tilde g}-1)},
&\mbox{$\frac{V}{2T_K}> e^{\delta}$}\cr
}
\label{eq:Gexpansion}
\ee
where $T_K$ is an energy scale set by the tunnel amplitude (the ``Kondo" scale) 
\be
T_K\propto |\Gamma|^{-1/
({\tilde g}^{-1}-1)}
\label{eq:TK}
\ee
and the coefficients $c_n$ are given by
\begin{equation}
c_n({\tilde g})=(-1)^{n-1}
\ \frac{\Gamma(n{\tilde g}+1)}{\Gamma(n+1)}
\ \frac{\Gamma(1/2)}{\Gamma(n({\tilde g}-1)+1/2)}\ .
\end{equation}
where $\Gamma(z)$ is the gamma function.
The domains of convergence of the dual series are restricted by
$\delta=[{\tilde g}\ln {\tilde g}+(1- {\tilde g})\ln (1-{\tilde
g})]/[2({\tilde g}-1)]$, where $\delta= \ln \gamma$.

It interesting to determine the asymptotic behavior of the differential
tunnel conductance $G_t$ for bot $V \gg T_K$ and $V \ll T_K$. In the
former (strong coupling) regime we fing that $G_t$ {\sl saturates} to
the value
\be
\lim_{V \gg T_K}G_t={\tilde g} {\frac{e^2}{h}}
\label{eq:saturation}
\ee
Notice that ${\tilde g}$ is {\sl not} equal to the filling factor $\nu$.
Hence, the large voltage differential tunnel conductance of a QPC is {\sl not equal}
to the quantum Hall conductance of the bulk FQH state!. In particular, 
for $\nu=1/3$, ${\tilde g}={\frac{1}{g'}}={\frac{1}{2}}$ and $G_t \to {\frac{1}{2}}
{\frac{e^2}{h}}$. 

Conversely, in the weak coupling regime, we get instead the well
known scaling behavior predicted by Wen~\cite{wen-tunnel}, and Kane and
Fisher~\cite{KF}
\be
\lim_{V \ll T_K}G_t={\tilde g} {\frac{e^2}{h}} c_1(\frac{1}{\tilde g})
\left({\frac{V}{2 T_K}}\right)^{2({\frac{1}{\tilde g}}-1)}
\label{eq:scaling}
\ee

\section{Tunneling Current, Conductance and Chang's Experiments}
\label{chang}

Hence, $G_t(V \gg T_K)$ is not the Hall conductance of the $\nu=1/m$
bulk FQH state the electrons are tunneling into. What is it? It is the
bulk Hall conductance of the problem in which charge $1$ particles are
tunneling between equivalent FQH states with effective filling factor
$\nu_{\rm eff}=g'$. (This is the problem that was solved by Fendley,
Ludwig and Saleur using the Thermodynamic Bethe Anzats~\cite{FLS}.)
These results for a quantum point contact strongly suggest that the
large voltage conductance depends on tha nature of the contact.

\begin{figure}[h!]
\vspace{.2cm}
\noindent
\hspace{1.5 in}
\epsfxsize=2.6in
\epsfbox{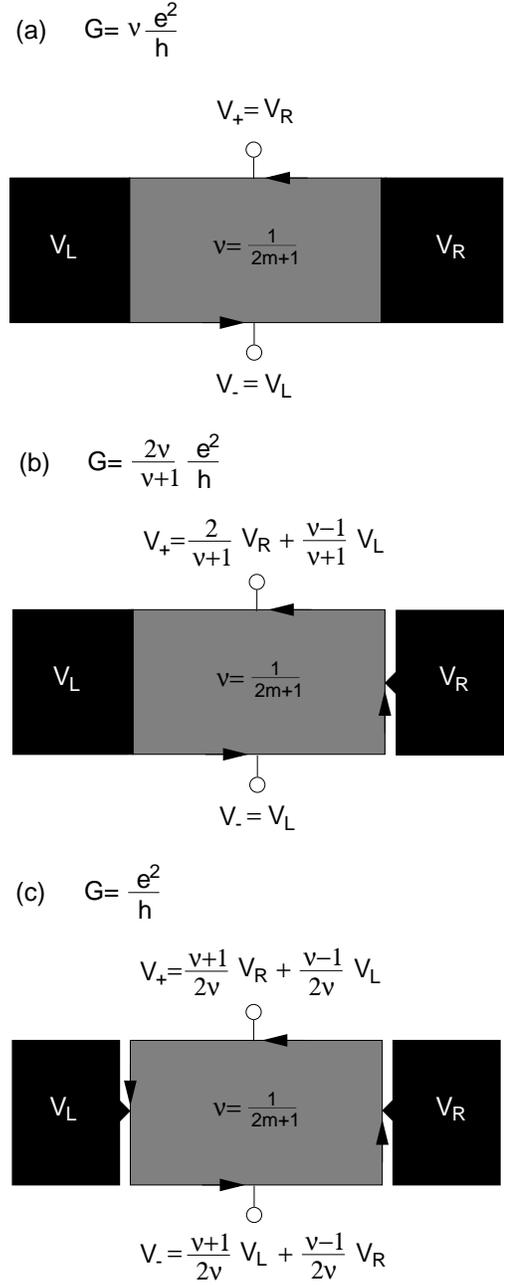}
\vspace{.5cm}
\caption{Different arrangements of contacts lead to different values of the 
large voltage differential conductance. }
\label{fig:cases}
\end{figure}

In view of this observation it is instructive to calculate the
asymptotic large voltage tunnel current and conductance for several cases (see
figure \ref{fig:cases}). 
\subsection{Uniform Tunneling}
Let us begin with case (a) of figure \ref{fig:cases},
in which the two reservoirs (source and sink) are equilibrated with
different parts of the edge of the FQH at filling factor $\nu=1/m$. In
this case, the tunneling current is equal to the quantized Hall current
of the device and the conductance is the bulk quantum Hall conductance.
We will show below how this limit is attained.
\subsection{One Point Contact}
This is the same case we considered before. In the strong tunneling
region the tunneling curreny $I_t$ is found to be proportional to the
voltage drop accros the junction
\be
I_t=G_t (V_R-V_{in})
\label{eq:Itlarge}
\ee
In this regime, $V \gg T_K$, Eq.\ \ref{eq:Gexpansion} predicts the behavior
\be
G_t \to {\frac{1}{g'}}{\frac{e^2}{h}} \left[1-\ldots\right]
\label{eq:Gtlarge-single}
\ee
where $g'^{-1}={\frac{1}{2}}(1+\nu^{-1})$. The reservoir to the left of
the $\nu=1/m$ FQH state is in equlibrium with the edge. Thus,
$V_{in}=V_L$. But, inside the FQH state the injected (tunnel) current
becomes the Hall current (which circulates around the edge). Thus
$V_+=V_{out}$ is
\be
V_{+}-V_{-}={\frac{I_t}{\nu e^2/h}}
\label{eq:Vout}
\ee
From what we conclude that 
\be
V_+-V_-={\frac{G_t}{\nu e^2/h}} (V_R-V_-)={\frac{2}{1+\nu}} (V_R-V_L)
\label{eq:deltaV}
\ee
is the Hall voltage, and $V_+$ is given by
\be
V_+={\frac{2}{1+\nu}} V_R + {\frac{\nu-1}{\nu+1}} V_L
\label{eq:V+}
\ee
These equations show that the single point contact is a simple
realization of the dc voltage transformer proposed by Chklovsky and
Halperin~\cite{CH}.

Hence, the two-terminal conductance of case (b) is
\be
{\rm Conductance}\equiv {\frac{I_t}{V_R-V_L}}={\frac{2\nu}{1+\nu}}
{\frac{e^2}{h}}
\label{eq:two}
\ee
For $\nu=1/3$ we find the saturation (maximum) value ${\frac{1}{2}}
{\frac{e^2}{h}}$ that we discussed above. 

\subsection{Two Point Contacts}

In case (c) we have a FQH droplet conected by two separate QPC's to two
independent reservoirs at voltages $V_R$ and $V_L$ respectively. there
is no equilibration going on in this case.  The same line of argument
used for case (b) now tells us that 
\begin{eqnarray}
V_+-V_-=&&{\frac{\nu}{g'}}(V_R-V_L)={\frac{2}{1+\nu}} (V_R-V_-)
\nonumber\\
V_--V_+=&&{\frac{1}{\nu g'}} (V_L-V_+)={\frac{2}{1+\nu}} (V_L-V_-)
\label{eq:deltaV2}
\end{eqnarray}
Hence,
\begin{eqnarray}
V_+=&&{\frac{1+\nu}{2\nu}}V_R+ {\frac{\nu-1}{2\nu}}V_L
\nonumber\\
V_-=&&{\frac{1+\nu}{2\nu}}V_L+ {\frac{\nu-1}{2\nu}}V_R
\label{eq:deltaV3}
\end{eqnarray}
Therefore,
\be
V_+-V_-={\frac{1}{\nu}} (V_R-V_L)
\label{eq:deltapm}
\ee
The tunneling current $I_t$ is found to be
\be
I_t={\frac{e^2}{h}}(V_+-V_-)
\label{eq:tcurrent2}
\ee
which implies that the two-terminal conductance is
\be
G={\frac{e^2}{h}}
\ee
instead of the FQH conductance. This result is a simple consecuence of
charge conservation. It is analogous to the statement that in a wire the
two-terminal conducatnce does not depend on the Luttinger
parameter~\cite{mike}.

\subsection{Multiple Contacts, Equlibration and Chang's Experiments}

\begin{figure}[h!]
\vspace{.2cm}
\noindent
\hspace{1.25 in}
\epsfxsize=3.0in
\epsfbox{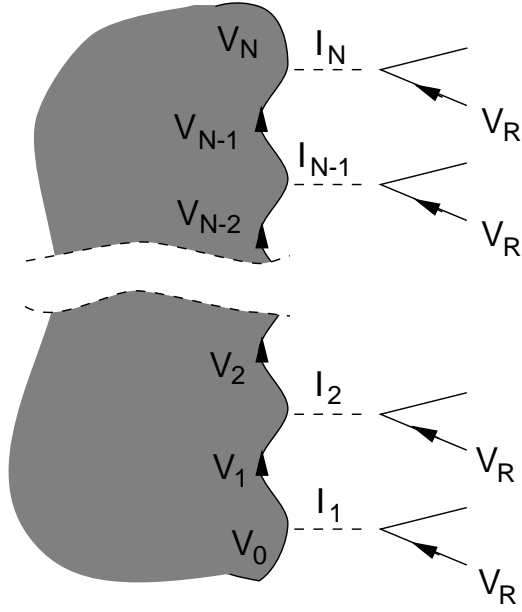}
\vspace{.5cm}
\caption{Tunnel for a broad equilibrated junction.}
\label{fig:tunnel-multi}
\end{figure}

Finally, let us consider case of figure \ref{fig:tunnel-multi}. 
In this case we have many QPC's. However, we will assume that the QPC's are 
sufficiently far apart  from each other that the reflected amplitudes
equilibrate with the source reservoir (the ``battery") and that as a
result there is no interference between QPC's. Notice, however, that the
FQH edge always remains coherent. Furthermore we will consider an arary of weak
tunnel junctions so that each junction remains in the perturbative regime.

In this case, which presumably applies to the actual experimental setup of
A.\ Chang and coworkers~\cite{chang}, the total
differential conductance $G_t$ at temperature $T$ and the total voltage drop 
$V=V_0-V_R$,  is given by~\cite{junction}
\begin{equation}
G=\nu\frac{e^2}{h}
\left\{
1-\frac{e^{-\frac{1}{2}\left(\frac{2\pi T}{T_K}\right)^{2(g-1)}}}
{\left[\frac{1}{\Gamma^2(g)}
\left(1-e^{-(g-1)\left(\frac{2\pi T}{T_K}\right)^{2(g-1)}}\right)
\left(\frac{V}{2\pi T}\right)^{2(g-1)}+1
\right]^{\frac{2g-1}{2(g-1)}}}\right\}\ .
\label{eq:fullotherg}
\end{equation}
where $g={\tilde g}^{-1}$. Here we have introduced the effective scale
$T_K^{eff}$
\be
\left(\frac{1}{T_K^{eff}}\right)^2=\frac{1}{3\nu} \sum_{n=1}^N
\left({\frac{1}{T_K^{(n)}}}\right)^2
\label{eq:Tkeff}
\ee
The expression for the differential tunnel conductance of 
Eq.\ \ref{eq:fullotherg} is used in reference \ref{ref:junction} (see figure
\ref{fig:chang} below) to fit the experimental data of A.\ Chang and
coworkers~\cite{chang}. 

At large voltages $V \gg
T_K^{eff}$, the differential tunnel conductance approaches the bulk FQH
conductance,
\be
\lim_{V \gg T_K^{rm}} G_t=\nu {\frac{e^2}{h}}
\label{eq:equlibration}
\ee
which shows that the edge and reservoir are in equlibrium. 
For $T\ll T \ll T_K$, the differential conductance of Eq.\ \ref{eq:fullotherg}
exhibits scaling behavior, $G_t \propto V^\alpha$, with an exponent
$\alpha=2g-1$. For $\nu=1/3$, we find $\alpha=3$.

\begin{figure}[h!]
\vspace{.2cm}
\noindent
\hspace{1.5 in}
\epsfxsize=4.0in
\epsfbox{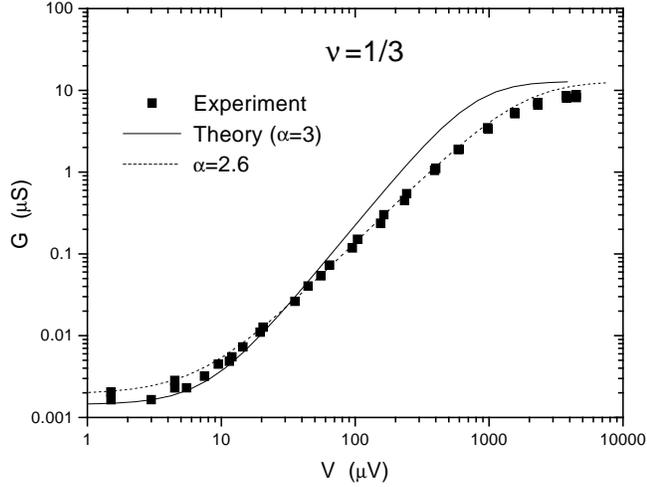}
\vspace{.5cm}
\caption{Comparison between experimental and theoretical differential 
conductances for
tunneling between an electron gas and a $\nu=1/3$ FQH state. For $T=25$ mK,
we get a best fit for $T_K=11$ mK and $\alpha=3$. The dotted line is
a fit with an effective exponent $\alpha=2.7$ and $T_K=26$ mK (see the text). 
(Experimental data courtesy of A.\ Chang.)}
\label{fig:chang}
\end{figure}

It is also possible to calculate the asymptotic conductance of a set of
$N$ junctions in the strong coupling limit provided they are
sufficiently far apart that there is no interference. For a set of $N_L$ and $N_R$ 
junctions between the FQH fluid and a left (L) reservoir, 
and a right (R) reservoirs, the result 
is~\cite{junction}
\be
G_{N_L,N_R}=
\frac{\left[1-\left(\frac{\nu-1}{\nu+1}\right)^{N_L}\right]
\left[1-\left(\frac{\nu-1}{\nu+1}\right)^{N_R}\right]}
{\left[1-\left(\frac{\nu-1}{\nu+1}\right)^{N_L+N_R}\right]}
\ \nu\frac{e^2}{h}
\label{eq:manycontacts}
\ee
Notice that these values of $G_t$ depend on both $N_L$ and $N_R$. 

\section{Strong coupling regime, non-Fermi liquid behavior and Andreev processes}
\label{andreev}

Finally, I want to discuss the physics of the strong coupling 
regime ($\Gamma \to \infty$ )of a single quantum point contact between
an Laughlin state and a Fermi liquid. This regime exhibits a number of very
interesting and fascinating behaviors. In the first part of this talk we
saw that the behavior of a single point contact is governed by the
non-trivial crossover energy scale $T_K$. Thus, we will be interested in
the behavior of the junction at $T=0$ and $V\gg T_K$. In what follows I
will assume that for voltages in this range, but smaller than a
natural cutoff energy scale $D < \hbar \omega_c$, where $\omega_c$ is
the cyclotron frequency. Here I will make the implicit assumption that
``more irrelvant" operators can be neglected in the description of the
junction. In principle such operators exist and their effects becomes large
 at strong coupling.  However, for any reasonable
model of the junction their coupling constants are small. Thus there 
should be a reasonably wide voltage range over which the effects of these
 operators can be ignored. Much of the discussion of this section is
based on the results of my work with Sandler and Chamon~\cite{andreev}.

Let us consider the strong coupling regime $\Gamma \to \infty$ of a
single junction. Alternatively, using duality, we can think of this
regime as ${\tilde \Gamma} \to 0$. we noted before that the strong
coupling fixed point is infrared unstable since the tunneling operator
(in the dual picture) $\cos ({\sqrt{2g'}}{\tilde \varphi}_-)$ has
(boundary) scaling dimension $g'=2/(1+m)<1$. (Here I am using the
``unfolded" picture and ${\tilde \varphi}_-$ is a chiral field.) In what
follows I will use the filling factor $\nu$ (instead of the denominator
$m$) to avoid notational confusions.
\begin{figure}[h!]
\vspace{.2cm}
\noindent
\hspace{1.5 in}
\epsfxsize=3.0in
\epsfbox{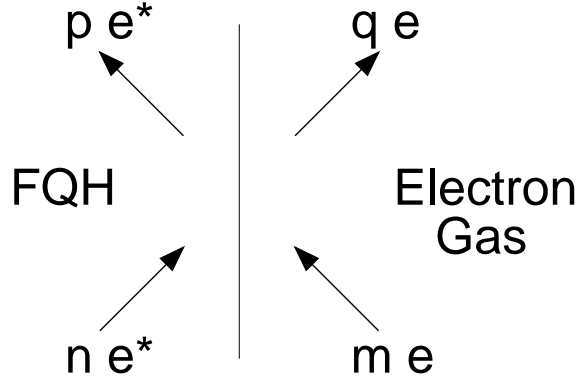}
\vspace{.5cm}
\caption{Scattering vertex for solitons and electrons.}
\label{fig:vertex}
\end{figure}

Let us consider the scattering process depicted in figure
\ref{fig:vertex}. In the incoming state we have $m$ electrons on the
Fermi liquid side (with total charge $me$), and $n$ quasiparticles on the FQH side
(with total charge $ne^*$). (Here $m$ is
an arbitary integer, unrelated to the denominator of $\nu$!.) Likewise, the
outgoing state has $q$ electrons (and charge $qe$) and $p$
quasiparticles (and charge $pe^*$). The $S$-matrix associated with this
process is contained in the correlation function
\begin{equation}
\langle :
e^{i\sqrt{\nu}p \phi_1^{out}} e^{iq \phi_2^{out}}::
e^{-i\sqrt{\nu} n \phi_1^{in}} e^{-i m \phi_2^{in}}:
\rangle
\label{eq:m1value}
\end{equation}
where I have used the original fields (unrotated and undualized!). In
the weak coupling fixed point $\Gamma=0$ this amplitude factorizes
(since the coupling constant is zero). The elementary scattering
processes at the weak coupling fixed point are elastic reflections of
electrons and quasiparticles at the junction (which behaves here as a
hard wall).

\begin{figure}[h!]
\vspace{.2cm}
\noindent
\hspace{.5 in}
\epsfxsize=4.0in
\epsfbox{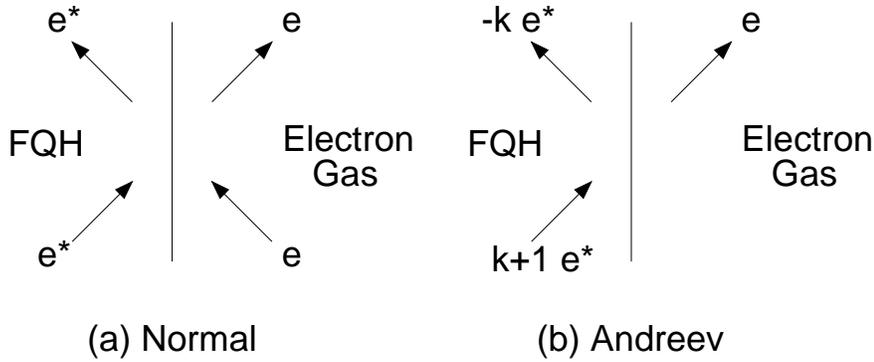}
\vspace{.5cm}
\caption{Elementary processes at strong coupling. Notice the Andreev process.}
\label{fig:elementary}
\end{figure}

The situation is drastically different at the strong coupling fixed
point. Not only there is no factorization but, instead of perfect
reflection, there are non-trivial {\sl selection rules} for the allowed
scattering processes. An elementary calculation~\cite{andreev} shows
that processes such as the simple reflection of a single electron at the
junction is forbidden: the amplitude vanishes exactly at ${\tilde
\Gamma}=0$. The same applies to the reflection of a single FQH
quasiparticle at the junction. In contrast, as shown in figure
\ref{fig:elementary}, the elementary allowed processes are of two types:
(a) $e-e^*$ scattering, that is, the {\sl scattering} of an electron off
a quasiparticle, without a charge exchanged across the junction, and (b)
the {\sl Andreev} process in which there are $k+1$ quasiparticles in the
initial state (for $\nu=1/(2k+1)$ and $k$ a positive integer) and the
final state has one transmitted electron and $k$ quasiholes. This
process is analogous to an Andreev reflection at a
normal-superconducting (NS) interface. (In the NS problem, the initial
state has one electron and the final state has a Cooper pair plus a
reflected hole.)

\begin{figure}[h!]
\vspace{.2cm}
\noindent
\hspace{1.5 in}
\epsfxsize=3.0in
\epsfbox{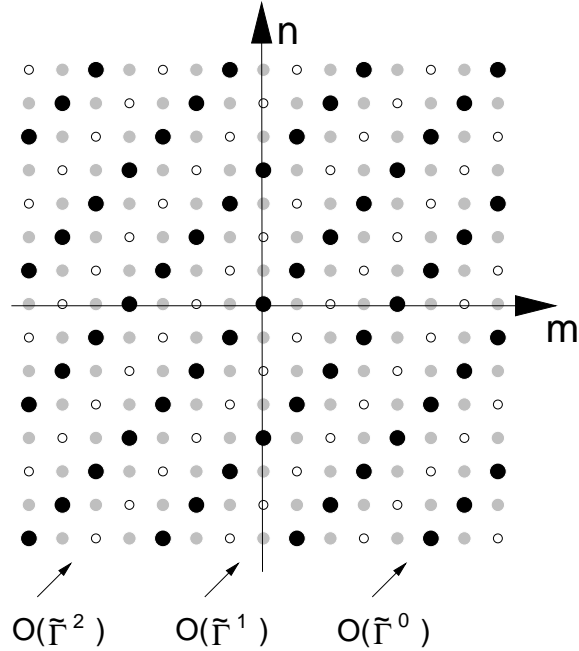}
\vspace{.5cm}
\caption{Lattice of scattering processes at different orders in 
the dual coupling constant.}
\label{fig:lattice}
\end{figure}

The selection rules exactly at ${\tilde \Gamma}=0$, 
for the case of a more general process 
of figure \ref{fig:vertex}, can be summarized by the matrix equation
\begin{equation}
\left(\matrix{q \cr p}\right)
=
{\bf M}
\left(\matrix{
m\cr n}\right)\ ,
\qquad
{\bf M} = \,
\left(
\matrix{
\frac{1 - \nu}{1 + \nu} & \; \frac{2 \nu}{1 + \nu}  \cr
\frac{2}{1 + \nu}       & \; -\frac{1 - \nu}{1 + \nu}}
\right)\ .
\end{equation}

Since $m, n, p, q$ are all integers, then it is easy to see that not all
combinations of integers are allowed since the entries of the matrix $M$ are 
fractional numbers. We can think of this equation as  a 
map from the lattice $(m,n)\in {\bf Z}_2$ onto itself (the points $(p,q)$).
The allowed processes at ${\tilde \Gamma}=0$ form the sublattice of
dark dots shown in figure \ref{fig:lattice} for the special case of
$\nu=1/7$. The sublattice of allowed processes is spanned by the vectors
${\vec a}_1=(1,1)$ (which represents elastic electron-quasiparticle
scattering) and ${\vec a}_2=(0,k+1)$ (the Andreev process).

For small $\tilde\Gamma$, 
the in and out quantum numbers are now related by
\begin{equation}
\left(
\matrix{q \cr p }
\right) = {\bf M}\,
\left(
\matrix{m \cr n}
\right)\ +\ l{\bf t}\ ,
\qquad
{\bf t}=\left( \matrix{ \frac{2\nu}{1+\nu} \cr -\frac{2}{1+\nu}}
\right)
\label{eq:inoutl}
\end{equation}
and $|l|$ is the order of the expansion
($\tilde\Gamma^{|l|}$). These higher order processes are shown as the shaded
dots on the lattice of figure \ref{fig:lattice}. For example, the
process in which there is just one lectron in the incoming state and
alos just one electron in the outgoing state ( a reflection), is
forbidden at ${\tilde \Gamma}=0$ (which is to say, the one-body $S$-matrix 
for electrons is zero), but it is non-zero at order ${\tilde \Gamma}$.
However, an explicit calculation shows that this amplitude has a branch
cut (in energy or momentum) instead of a pole as in weak coupling.
Hence it is no longer possible to give a
particle interpretation to this process since there is no
single particle scattering but instead only multiparticle processes: the
leading contribution has a broad spectrum and looks like ``incoherent"
scattering. 

Moreover, it is possible to show that the junction has a non-zero finite
entropy at the strong coupling fixed point. In fact, this phenomenon is
well know to in the context of the effective theory of the junction, the
boundary sine-Gordon problem. Boundary sine-Gordon is a (boundary)
conformal field theory (CFT). As we noted above, the effect of the tunneling 
operator is to induce a flow of boundary conditions. This is a very
general phenomenon in boundary CFT's, studied in great detail and
generality by Affleck and Ludwig~\cite{flow}. They conjectured that in all
boundary CFT's there
exists a boundary degeneracy (or entropy) which flows (under ${\tilde
\Gamma}$ in our case) much as the central (Virasoro) charge flows in the
bulk CFT. In particular Fendley, Saleur and Warner~\cite{FSW} used the
Thermodynamic Bethe Anstaz and found that,
in the thermodynamic limit, the total entropy flows frome
the non-extensive but finite value 
\be
S(0)= \lim_{T \to 0} \left(-{\frac{F}{T}}\right)={\frac{1}{2}}
\ln (k+1)
\label{eq:entropy}
\ee
at ${\tilde \Gamma}=0$, to the value $S(0)=0$ at $\Gamma=0$. Here $F$ is
the total Free energy of the boundary sine-Gordon theory. In fact, for
the special case of a junction of a $\nu=1/3$ FQH fluid and a Fermi
liquid, $\nu=1$, we find $S(0)={\frac{1}{2}}\ln 2$, which is the
boundary entropy of the two-channel Kondo problem!

Behaviors of these sort, both in the $S$-matrix and in the entropy,
have been found previously in the
multi-channel Kondo problem, which has a ``non-Fermi liquid" fixed point.
Clearly the physics of the strongly coupled junction is very similar.

\section*{Acknowledgments}

I am deeply indebted to my collaborators  Claudio Chamon and Nancy Sandler for
their insights. I am also very grateful to Albert Chang for many
discussions and for making his data available to us. 
I thank the organizers of this  
XXXIVth Rencontres de Moriond, Condensed Matter Physics Meeting {\sl 
Quantum Physics at the Mesoscopic Scale} for having arranged this great meeting. 
I am particularly grateful to Dr.\ Christian Glattli for his kind 
invitation to this great workshop. This work was supported in part
by the NSF grant number DMR98-17941 at the University of Illinois at
Urbana-Champaign and by the John Simon Guggenheim Memorial Fellowship
Foundation.

\end{document}